\documentclass[pra,twocolumn,amsmath,amssymb,nofootinbib]{revtex4}

\usepackage{graphicx}

\begin{document}
\title{NPA Hierarchy and Extremal Criterion in the Simplest Bell Scenario}
\author{Satoshi Ishizaka}
\affiliation{Graduate School of Advanced Sciences and Engineering,
Hiroshima University,
1-7-1 Kagamiyama, Higashi-Hiroshima, 739-8521, Japan}
%
\begin{abstract}
It is difficult to establish an analytical criterion to identify the
boundaries of quantum correlations, even for the
simplest Bell scenario. Here, we briefly reviewed the plausible analytical criterion,
and we found a way to confirm the extremal conditions from another direction.
For that purpose, we analyzed the Navascu\'es-Pironio-Ac\'{\i}n (NPA)
hierarchy to study the algebraic structure and found that 
the problem could not be simplified using $1\!+\!AB$ level. 
However, considering the plausible
criterion, the $1\!+\!AB$ and second levels for correlations were equal, and the
extremal condition in the simplest Bell scenario was replaced by that
in the $1\!+\!AB$ level. Thus, the correctness of the
plausible criterion was verified, and the results demonstrated that the plausible
criterion held, thereby explaining its simplicity. It seemed plausible, but now 
it becomes more certain.
\end{abstract}
%
\maketitle
%

\section{Introduction}

In 1964, Bell showed that nonlocal correlations predicted by quantum
mechanics were inconsistent with local realism \cite{Bell64a}. 
Although the non-local correlations did not contradict the no signaling principle, the strength of the
quantum correlations was more restricted than that allowed by the no-signaling
principle \cite{Tsirelson80a,Popescu94a}. Since then, fundamental principles
have been investigated to address this discrepancy \cite{Brunner14a,Popescu14a,Oas16a}.
Furthermore, the origin of fundamental
principles, such as uncertainty and nonlocality, has been investigated \cite{Oppenheim10a,Ramanathan18a}.
However, in these studies, clarifying the boundaries of
quantum correlations was difficult. Indeed, an analytical criterion for 
identifying the boundaries has not yet been established, even for the simplest Bell scenario.

In the simplest Bell scenario, Tsirelson showed that the Bell inequality of the
Clauser-Horne-Shimony-Holt (CHSH) type \cite{Clauser69a} is violated up to
$2\sqrt{2}$ by quantum correlations \cite{Tsirelson80a}. The correlation
that attains the Tsirelson bound is an extremal point in a convex set of quantum
correlations. When the marginal probabilities of obtaining the measurement results
are unbiased (zero-marginal case), the boundaries are identified using the
analytical criterion Tsirelson-Landau-Masanes (TLM)
\cite{Tsirelson87a,Landau88a,Masanes03a}.
Recent studies have shown that there are many characteristics of the boundaries in a
quantum set \cite{Le23a}.

In the general case where the marginals may be biased (full-marginal case), obtaining
the analytical criterion is a long-standing open problem. In fact, only a few
examples of analytical solutions are known   
\cite{Acin12a,Rabelo12a,Wolfe12a,Ramanathan18a,Goh18a}.
The geometry of a quantum set exhibits rich counterintuitive
features \cite{Goh18a}. We propose a plausible analytical criterion to identify all extremal points
(a point is an extremal point if and only if the equalities Eq.\ (\ref{eq: Condition 1}) and
(\ref{eq: Condition 2}) are simultaneously satisfied) \cite{Ishizaka18a}.
Recently, some
quantum states have been discussed, and it has been shown that the plausible criterion
is a strong indication, which, however, remains unproven \cite{Mikos23a}.

The plausible criterion proposed in \cite{Ishizaka18a} is simpler
than expected. Therefore, we analyze the Navascu\'es-Pironio-Ac\'{\i}n
(NPA) hierarchy \cite{Navascues07a,Navascues08a,Navascues15c} to study the algebraic
structure. 
Specifically, the sets of correlations are denoted 
$Q^{(1)}$, $Q^{(1\!+\!AB)}$, $Q^{(2)}$, $\cdots$ (it is also called the correlation levels
between the first, $1\!+\!AB$, second, and others correspondingly).
By definition, the inclusion relation generally
$Q^{(1)} \supseteq Q^{(1\!+\!AB)} \supseteq Q^{(2)}\supseteq\cdots\supseteq Q$,
where $Q$ is the set of correlations in the simplest Bell scenario
and all the extremal points are conjectured by the plausible criterion. 
If a phenomenon (in a Bell inequality) is explained up to the $1\!+\!AB$
level, then $Q^{(1)} \supseteq Q^{(1\!+\!AB)} = Q^{(2)}=\cdots=Q$ holds true.
This becomes clear upon examining the inclusion relationship between the $1\!+\!AB$ and
second levels (at this time the following condition is automatically satisfied by a rank loop:
$Q^{(2)}=Q^{(3)}=\cdots$ \cite{Navascues08a}).
In the NPA analysis, the simplicity of the plausible criterion is persuasive, especially when 
the $1\!+\!AB$ and second levels are equal (just as the TLM criterion is explained by the first
and second levels). 
Otherwise, in this study, we say nothing exceptional about $1\!+\!AB$ level.

In this paper, we briefly review the plausible criterion for convenience,
and we find a way to confirm the extremal conditions from another direction.
For that purpose, we analyze the NPA hierarchy to study the algebraic
structure, and the result show that the $1\!+\!AB$ level is not exceptional.
After all, the problem cannot be simplified using the $1\!+\!AB$ level.
However, as far as the plausible criterion is concerned, 
the $1\!+\!AB$ and second levels for correlations are equal, and the extremal
condition in the simplest Bell
scenario is replaced by that in the $1\!+\!AB$ level. Thus, the correctness 
of the plausible criterion is clarified,
and the results demonstrate that it holds true, which explains its simplicity.

\section{Plausible criterion}

We now clarify the plausible criterion obtained in the
two parity Alice and Bob' system; the projective measurements of rank 1 are performed
in a two-qubit entangled state.
Specifically, without loss of generality, the observables are
\begin{equation}
A_x=\cos \theta^{A}_x \sigma_1 + \sin \theta^{A}_x \sigma_3,\hbox{~}
B_y=\cos \theta^{B}_y \sigma_1 + \sin \theta^{B}_y \sigma_3,
\label{eq: Parameterization 1}
\end{equation} 
where $(\sigma_1,\sigma_2,\sigma_3)$ are the Pauli matrices, and the two-qubit entangled
state is
\begin{equation}
|\psi\rangle=\cos\chi|00\rangle+\sin\chi|11\rangle
\hbox{~~~($0\!<\!\chi\!\le\!\pi/4$)}.
\label{eq: Psi}
\end{equation}
Based on this parameterization, we obtain
\begin{eqnarray}
\langle A_xB_y\rangle&=& \sin\theta^{A}_x\sin\theta^{B}_y+\cos\theta^{A}_x\cos\theta^{B}_y \sin2\chi,
\label{eq: Cxy}   \\
\lefteqn{\hspace{-10mm}\langle A_{x}\rangle= \sin\theta^{A}_x \cos2\chi,
\hbox{ }   
\langle B_{y}\rangle= \sin\theta^{B}_y \cos2\chi.} \label{eq: By}
\end{eqnarray}

The plausible criterion is classified into two categories. The first is 
based on the limitations of entangled states. Based on the analysis presented in Ref. \cite{Ishizaka18a}, 
the usage of the entanglements are specified by the eight parameters $S^{\pm}_{xy}$,
which obey the following correlations:
\begin{eqnarray}
S^{\pm}_{xy}&\equiv&\frac{1}{2}\left[J_{xy}
\pm\sqrt{J^{2}_{xy}-4K^{2}_{xy}}\right], \\
J_{xy}&\equiv&\langle A_xB_y\rangle^2-\langle A_x\rangle^2-\langle B_y\rangle^2+1, \\
K_{xy}&\equiv&\langle A_xB_y\rangle-\langle A_x\rangle\langle B_y\rangle.
\end{eqnarray}
Here, for each $x$ and $y$,  $\sin2\chi\!=\!S^+_{xy}$ or $\sin2\chi\!=\!S^-_{xy}$
always holds true. If a two-qubit realization cannot be guaranteed for Eq.\ (\ref{eq: Cxy}),
the condition
\begin{equation}
\prod_{xy}\{(1-S^{+}_{xy})\langle A_x B_y\rangle-\langle A_x\rangle\langle B_y\rangle\}\ge 0,
\label{eq: positivity}
\end{equation}
is necessary; see the Supplemental Material of Ref. \cite{Ishizaka18a}. The maximum use of
the entanglements is $S^+_{00}\!=\!S^+_{01}\!=\!S^+_{10}\!=\!S^+_{11}$.

The second category is the cryptographic quantum bound \cite{Ishizaka17a}. The TLM
inequality becomes an equality whose correlation is divided guessing probability.
The scaled TLM inequality is expressed as:
\begin{eqnarray}
\left|\tilde C_{00}\tilde C_{01}-\tilde C_{10}\tilde C_{11}\right|
&\le& (1-\tilde C^{2}_{00})^{1/2}
(1-\tilde C^{2}_{01})^{1/2} \cr
&&+(1-\tilde C^{2}_{10})^{1/2}(1-\tilde C^{2}_{11})^{1/2}
\end{eqnarray}
where $\tilde C_{xy}\!=\!\langle A_xB_y\rangle/D^{B}_x$ and 
$\tilde C_{xy}\!=\!\langle A_xB_y\rangle/D^{A}_y$.
The guessing probabilities are as follows:
\begin{eqnarray}
D^{B}_x &\!\!\!=\!\!\!&\sqrt{\sin^2 \theta^{A}_x+\cos^2 \theta^{A}_x\sin^2 2\chi}
=\hbox{tr}|\rho^{B}_{1|x}\!-\!\rho^{B}_{-1|x}|, \\
D^{A}_y &\!\!\!=\!\!\!&\sqrt{\sin^2 \theta^{B}_y+\cos^2 \theta^{B}_y\sin^2 2\chi}
=\hbox{tr}|\rho^{A}_{1|y}\!-\!\rho^{A}_{-1|y}|,
\end{eqnarray}
where
$\rho^{B}_{a|x}\!=\!\hbox{tr}_A\frac{I+aA_x}{2}|\psi\rangle\langle\psi|$
and
$\rho^{A}_{b|y}\!=\!\hbox{tr}_B\frac{I+bB_y}{2}|\psi\rangle\langle\psi|$.
The NPA inequality is also extended to include $D^{B}_x$ and $D^{A}_y$
(see Eq.\ (B4) in Ref.\ \cite{Ishizaka17a}).
However, it is not used because full-rank is necessary and
a two-qubit pure state does not satisfy the equality. Therefore, it is not an
appropriate extremal condition.

Thus, our plausible criterion is as follows:
\begin{eqnarray}
\lefteqn{\hspace{-20mm}S^+_{00}=S^+_{01}=S^+_{10}=S^+_{11},} \label{eq: Condition 1} \\
\left|\tilde C_{00}\tilde C_{01}-\tilde C_{10}\tilde C_{11}\right|
&\!\!=\!\!& (1-\tilde C^{2}_{00})^{1/2} (1-\tilde C^{2}_{01})^{1/2} \nonumber \\
&&+(1-\tilde C^{2}_{10})^{1/2}(1-\tilde C^{2}_{11})^{1/2} \label{eq: Condition 2}.
\end{eqnarray}
A point is an extremal point if and only if the five equalities are simultaneously satisfied.
The scaled TLM condition must be satisfied in both 
$\tilde C_{xy}\!=\!\langle A_xB_y\rangle/D^{B}_x$ and 
$\tilde C_{xy}\!=\!\langle A_xB_y\rangle/D^{A}_y$, and there are two equalities.

\section{NPA hierarchy and simplest Bell scenario}

\begin{table}[t]
\begin{tabular}{cccc}
x & QB$_2^{(8)}$ & $1\!+\!AB$ & 2nd \\
\hline
0.0 & 2.82842712474619 & 2.82842712474619 & 2.82842712474619 \\
0.4 & 2.88444102037119 & 2.88444102037119 & 2.88444102037119 \\
0.8 & 3.04630924234556 & 3.04630924234556 & 3.04630924234556 \\
1.2 & 3.29848450049413 & 3.29848450049413 & 3.29848450049413 \\
1.6 & 3.62215405525497 & 3.62215405525497 & 3.62215405525497 \\
2.0 & 4.00000000000000 & 4.00000000000000 & 4.00000000000000 \\
\hline
\\
x & QB$_3^{(8)}$ & $1\!+\!AB$ & 2nd \\
\hline
0.0 & 2.82842712474619 & 2.82842712474619 & 2.82842712474619 \\
0.2 & 2.83091685885720 & 2.83091685885720 & 2.83091685885720 \\
0.4 & 2.83923308963559 & 2.83923308963559 & 2.83923308963559 \\
0.6 & 2.85676984164748 & 2.85676984164748 & 2.85676984164748 \\
0.8 & 2.89417689813970 & 2.90075597099059 & 2.89417689813970 \\
1.0 & 3.00000000000000 & 3.01789221335227 & 3.00737232088269 \\
1.2 & 3.20000000000000 & 3.20000000000000 & 3.20000000000000 \\
1.4 & 3.40000000000000 & 3.40000000000000 & 3.40000000000000 \\
1.6 & 3.60000000000000 & 3.60000000000000 & 3.60000000000000 \\
1.8 & 3.80000000000000 & 3.80000000000000 & 3.80000000000000 \\
2.0 & 4.00000000000000 & 4.00000000000000 & 4.00000000000000 \\
\hline
\end{tabular}
\caption{
Correlations of QB$_2^{(8)}$ and QB$_3^{(8)}$, obtained using the 
quantifier elimination algorithm in Ref.\ \cite{Wolfe12a}.
The superscript ${(8)}$ denotes a fully-marginal 8D case to distinguish
the zero-marginal 4D case.
The quantum values (theoretical values) for QB$_2^{(8)}$ are 
$\sqrt{2x^2+8}$, and for QB$_3^{(8)}$ are $\frac{\sqrt{(2-x^2)(4-3x^2)}-x^2}{1-x^2}$ ($x\!\le\!1$) and
$x+2$ ($1\!\le\!x\!\le\!2$), respectively.
The calculated NPA $1\!+\!AB$ and 2nd levels are also listed. 
}
\label{tbl: QB}
\end{table}

We now analyze the NPA hierarchy 
and use explicit examples to determine whether $1\!+\!AB$ level is exceptional in the simplest Bell scenario.
Table \ref{tbl: QB} lists the correlations of the quantum bounds QB$_2^{(8)}$ and QB$_3^{(8)}$, which
are obtained using the quantifier elimination algorithm \cite{Wolfe12a}. 
QB$_3^{(8)}$is discussed in 
relation to the plausible criterion \cite{Mikos23a}.
In the table, the measures
$x\langle A_{0}\rangle
+\langle A_0B_0\rangle+\langle A_1B_0\rangle+\langle A_0B_1\rangle
-\langle A_1B_1\rangle$ for QB$_2^{(8)}$ and 
$x\langle A_{0}\rangle\!+\!x\langle A_{1}\rangle\!-x\!\langle B_{0}\rangle
+\langle A_0B_0\rangle+\langle A_1B_0\rangle+\langle A_0B_1\rangle
-\langle A_1B_1\rangle$ for QB$_3^{(8)}$ are maximized,
and the quantum values (theoretical values) are provided in figure
caption \cite{Wolfe12a}.
The calculated NPA $1\!+\!AB$ and 2nd levels are also listed,
where a highly accurate 
multiple-precision semidefinite programming solver is used \cite{SDPA},
and the same measures as the quantum values are maximized.
In QB$_2^{(8)}$, the $1\!+\!AB$ calculations are consistent with the quantum values
 (within the limits of numerical accuracy). In QB$_3^{(8)}$, the $1\!+\!AB$ calculations
deviate from the quantum values of approximately  
$0.8\!\le\!x\!\le\!1.0$.
Here, the endpoint $x\!=\!1.0$ is the classical regime according to the
TLM criterion, but is influenced by the quantum regime with instability. 
Specifically, the deviation rapidly converges and remains within $x\!=\!0.698$
(figure is not shown); $x\!\le\!0.7$ seems accurate throughout corresponding to
the non-deviate correlation. 
Thus, we conclude that deviate and non-deviate correlations coexist,
and that there is nothing exceptional about $1\!+\!AB$ level.
The certificate is QB$_3^{(8)}$ with $0.698\!\le\!x\!\le\!1$.

\begin{figure}[t]
\centerline{\scalebox{0.5}[0.5]{\includegraphics{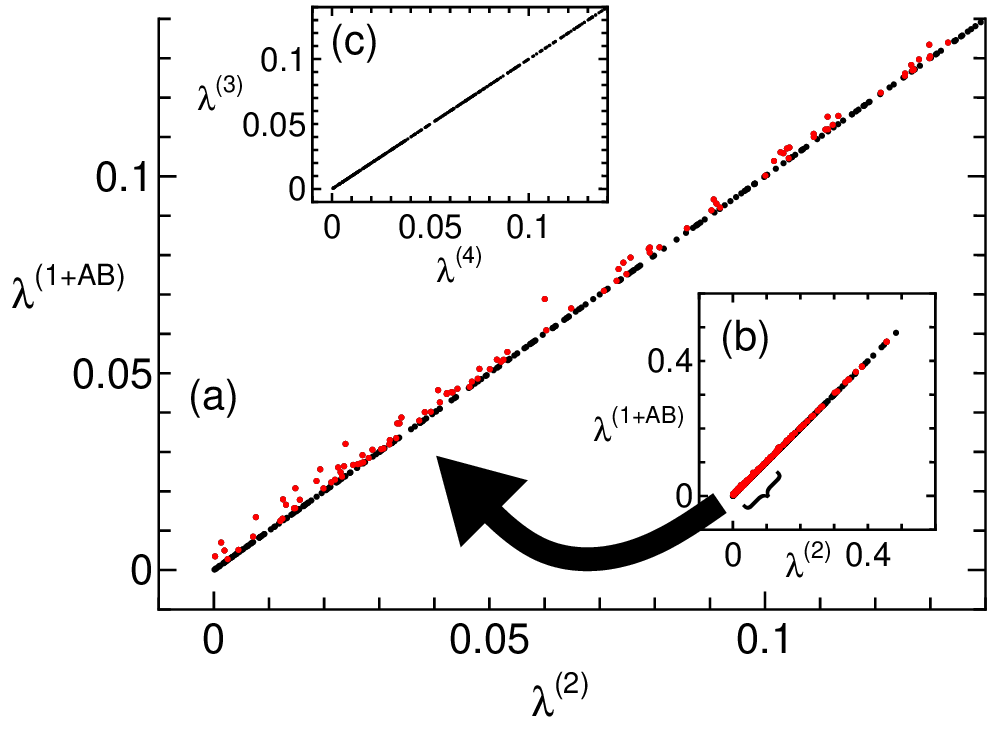}}}
\caption{
Maximum of $\lambda$, where $\Gamma\!-\!\lambda\openone\!\ge\!0$ at
NPA $1\!+\!AB$ level (vertical axis) and 2nd level (horizontal axis), obtained using the
semidefinite programming solver \cite{SDPA}. (a) Plotted area near the origin and
(b) full scale. The deviations from the straight line are indicated in red. 
(c) With data at 3rd (vertical) and 4th (horizontal) levels, for the levels equal to or greater
than this, the two data are on a straight line.
The number of data points are 500 to distinguish each data point. 
}
\label{fig: NPA}
\end{figure}

Another semidefinite program demonstrated an inclusion relationship between
NPA $1\!+\!AB$ and second levels. It determines the maximum of $\lambda$ such that  
$\Gamma\!-\!\lambda I\!\ge\!0$ 
($\Gamma\!-\!\lambda I$ is nonlocal), 
where $\Gamma$ is the certificate matrix \cite{Navascues08a}.
Here, random initial points $\{\langle A_xB_y\rangle,\langle A_x\rangle,\langle B_y\rangle\}$,
which are included in $\Gamma$,
are common for the $1\!+\!AB$ and second levels. However,  
the maximum $\lambda$ calculated using $1\!+\!AB$ and second denotes $\lambda^{(1+AB)}$
and $\lambda^{(2)}$, respectively.
Figure \ref{fig: NPA} (a) and (b) shows the results of $\lambda^{(1+AB)}$ (y-axis) and
$\lambda^{(2)}$ (x-axis). Because $Q^{(1+AB)}\supseteq Q^{(2)}$, $\lambda^{(1+AB)}$ is the
same as $\lambda^{(2)}$ or shifted up.
When $\lambda^{(1+AB)}$ and $\lambda^{(2)}$ are on a
straight line, the $1\!+\!AB$ and second levels are always equal, 
and the correlations can be explained up to the $1\!+\!AB$ level.
However, in the figure, approximately 30\% of the data points deviate from
the straight line (indicated in red) and deviated and non-deviated correlations coexist.
Similarly, nothing exceptional is observed about the $1\!+\!AB$ level.
How many orders of magnitude do we need to consider?
Figure \ref{fig: NPA} (c) shows the data of $\lambda^{(3)}$ (y-axis) and
$\lambda^{(4)}$ (x-axis). At the levels equal to or greater than this, the two data are
on a straight line, and two levels such as $\lambda^{(3)}$ and $\lambda^{(4)}$ are always equal.
Thus, the second level is not enough (figure is not shown) and the third level is required.

\begin{figure}[tb]
\centerline{\scalebox{0.5}[0.5]{\includegraphics{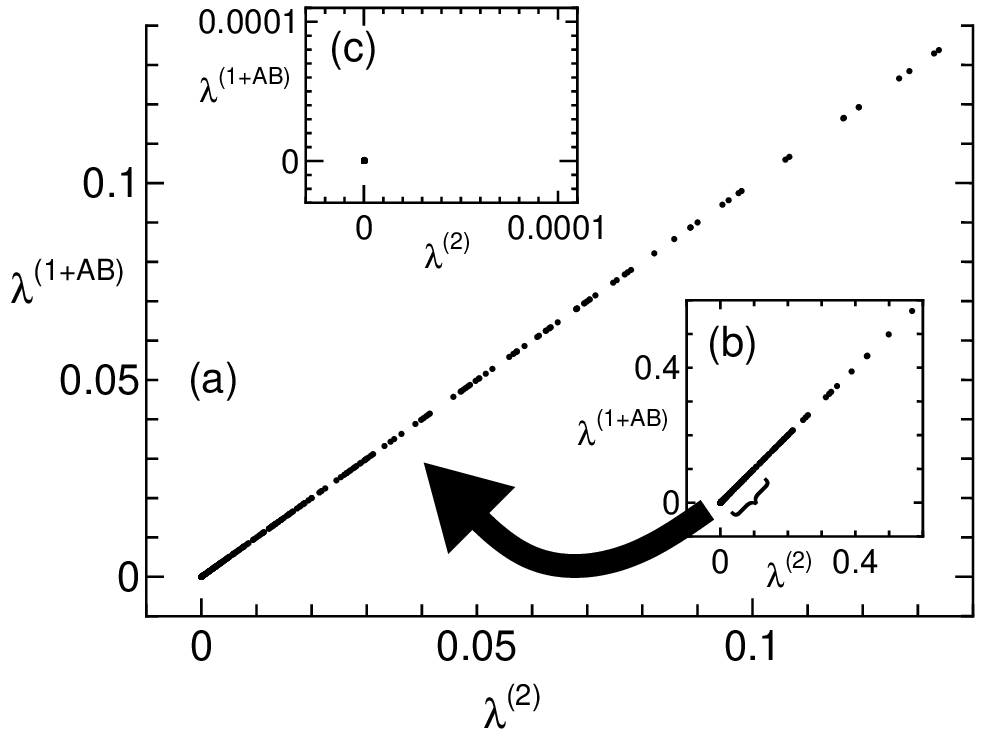}}}
\caption{
Two-qubit realizations and Eq.\ (\ref{eq: Condition 1})
are randomly specified as initial points. These graphs are similar to Fig.\ \ref{fig: NPA};
however, selected by initial points. (a) Plots showing the  
area near the origin and (b) full scale. All the data points are non-deviate from the straight line,
and the NPA $1\!+\!AB$ and 2nd levels are always equal.
(c) The plausible criteria (Eq.\ (\ref{eq: Condition 1}) and (\ref{eq: Condition 2}))
are fully specified, and all the data points converge to a single point (0,0).
}
\label{fig: NPA2}
\end{figure}

Two maximization problems exist: maximizing the quantum value (which depends
on the state to be maximized) and maximizing $\lambda$ such that
$\Gamma\!-\!\lambda I\!\ge\!0$.
Both yield the same results, and no exceptional results are observed for $1\!+\!AB$ level.
Thus, using the $1\!+\!AB$ level is unlikely to simplify the problem.
However, maximizing $\lambda$ such that $\Gamma\!-\!\lambda I\!\ge\!0$ is 
an interesting feature, which is revealed by selecting a fixed initial point in Fig.\ \ref{fig: NPA}.
Specifically, two-qubit realizations are randomly specified as the initial points 
(pure states Eqs.\ (\ref{eq: Psi})-(\ref{eq: By}) are specified considering Eq.\ (\ref{eq: positivity})).
Thus, mixed states are exclusively removed; however, this does not change significantly from Fig.\ \ref{fig: NPA}.
Thus, Eq.\ (\ref{eq: Condition 1}) is additionally specified
as the initial points. Figure \ref{fig: NPA2} (a)
shows the results where none of the data points deviate from the straight line. 
Therefore, because it is a pure state and satisfies 
Eq.\ (\ref{eq: Condition 1}), the $1\!+\!AB$ and second levels are equal.
This will further equalize the correlations of the simplest Bell scenario,
because a rank loop is established at the third level as shown Fig.\ \ref{fig: NPA} (c), 
but for these initial points the $1\!+\!AB$ and third levels are equal (figure is not shown).
\footnote{Because QB$_3^{(8)}$ is a pure state and satisfies
Eq.\ (\ref{eq: Condition 1}), the $1\!+\!AB$ and 
second levels are equal (QB$_3^{(8)}$ satisfies Eq.\ (\ref{eq: Condition 2}) \cite{Mikos23a},
because QB$_3^{(8)}$ is a solution to the plausible criterion).
However, in QB$_3^{(8)}$ in Table \ref{tbl: QB}, 
these levels are unequal, 
apparently because what is maximized is obviously different. Thus, 
even the relationship between the correlations varies depending on what you maximize
}{
Thus, when it is a pure state and Eq.\ (\ref{eq: Condition 1}) is satisfied,
the $1\!+\!AB$ and second levels are equal to those in the
simplest Bell scenario}.

We now consider what happens if the remaining plausible criteria are added to
Eq.\ (\ref{eq: Condition 2}).
Figure \ref{fig: NPA2} (c) shows the results where all the data points
($\lambda^{(2)}$,$\lambda^{(1+AB)}$) converge to a single point (0,0).
This convergence is explained as follows. 
First, the $1\!+\!AB$ and second levels are equal to the correlations
of the simplest Bell scenario because they satisfy Eq.\ (\ref{eq: Condition 1}).
Consequently, the extremal condition in the simplest Bell scenario is replaced by that 
in the $1\!+\!AB$ level.
Second, because $\lambda^{(1+AB)}$ and $\lambda^{(2)}$ are already optimal solutions
of $\Gamma\!-\!\lambda I\!\ge\!0$, we obtain
($\lambda^{(2)}$,$\lambda^{(1+AB)}$)=(0,0).
Therefore, after slight consideration, we find that the plausible criterion
must be extremal points of the $1\!+\!AB$ level to ensure that 
($\lambda^{(2)}$,$\lambda^{(1+AB)}$) converges to (0,0).
Conversely, if the plausible criterion is incorrect and not a flat boundary
point, convergence to $(0,0)$ does not occur. 
Thus, the correctness of the plausible criterion is verified,
although we are unsure of the existence of unnecessary flat boundary points.
The convergence to a single point proves that the plausible criterion holds.
Furthermore, the plausible criterion is simple because the simplest Bell scenario is
explained by the $1\!+\!AB$ level. 
However, the method described is suitable for obtaining predictions;
if predictions cannot be obtained, the method is unsuitable.
In such cases, the Monte Carlo method may be used. However, an event is unlikely to  
survive hundreds of thousands of Monte Carlo
trials. Our calculations do not reveal any exceptions to
Eq.\ (\ref{eq: Condition 1}) and (\ref{eq: Condition 2}) \cite{Ishizaka18a}.

\section{Summary}

In summary, we analyze the NPA hierarchy to study the algebraic structure in the simplest
Bell scenario.
The result is that the problem could not be simplified using $1\!+\!AB$ level. 
However, when the problem is limited to the plausible criterion,
the $1\!+\!AB$ and second levels for correlations are equal, and
the extremal condition in the simplest Bell scenario is replaced by that 
in the $1\!+\!AB$ level.
Although the existence of unnecessary flat boundary points remains unknown,
the correctness of the plausible criterion is ascertained, and the results
numerically demonstrate that the plausible criterion becomes more certain.
Furthermore, this explains the simplicity of the plausible criterion.

\acknowledgments{This work was supported by JSPS KAKENHI Grants No. 21K03389, 22K03452.}

\end{document}